\documentclass[longauth]{aa} 
\usepackage{graphicx}
\usepackage{lscape}
\begin{document}
   \title{HESS Observations and VLT Spectroscopy of PG\,1553+113}


\author{F. Aharonian\inst{1,13}
 \and A.G.~Akhperjanian \inst{2}
 \and U.~Barres de Almeida \inst{8} \thanks{supported by CAPES Foundation, Ministry of Education of Brazil}
 \and A.R.~Bazer-Bachi \inst{3}
 \and B.~Behera \inst{14}
 \and M.~Beilicke \inst{4}
 \and W.~Benbow \inst{1}
 \and K.~Bernl\"ohr \inst{1,5}
 \and C.~Boisson \inst{6}
 \and O.~Bolz \inst{1}
% ...
}

\offprints{Wystan.Benbow@mpi-hd.mpg.de 
or Catherine.Boisson@obspm.fr}

\institute{
Max-Planck-Institut f\"ur Kernphysik, P.O. Box 103980, D 69029
Heidelberg, Germany
\and
 Yerevan Physics Institute, 2 Alikhanian Brothers St., 375036 Yerevan,
Armenia
\and
Centre d'Etude Spatiale des Rayonnements, CNRS/UPS, 9 av. du Colonel Roche, BP
4346, F-31029 Toulouse Cedex 4, France
\and
Universit\"at Hamburg, Institut f\"ur Experimentalphysik, Luruper Chaussee
149, D 22761 Hamburg, Germany
\and
Institut f\"ur Physik, Humboldt-Universit\"at zu Berlin, Newtonstr. 15,
D 12489 Berlin, Germany
\and
LUTH, Observatoire de Paris, CNRS, Universit\'e Paris Diderot, 5 Place Jules Janssen, 92190 Meudon, 
France
\and
DAPNIA/DSM/CEA, CE Saclay, F-91191
Gif-sur-Yvette, Cedex, France
\and
University of Durham, Department of Physics, South Road, Durham DH1 3LE,
U.K.
\and
Unit for Space Physics, North-West University, Potchefstroom 2520,
    South Africa
\and
Laboratoire Leprince-Ringuet, Ecole Polytechnique, CNRS/IN2P3,
 F-91128 Palaiseau, France
\and 
Laboratoire d'Annecy-le-Vieux de Physique des Particules, CNRS/IN2P3,
9 Chemin de Bellevue - BP 110 F-74941 Annecy-le-Vieux Cedex, France
\and
APC, 11 Place Marcelin Berthelot, F-75231 Paris Cedex 05, France 
\thanks{UMR 7164 (CNRS, Universit\'e Paris VII, CEA, Observatoire de Paris)}
\and
Dublin Institute for Advanced Studies, 5 Merrion Square, Dublin 2,
Ireland
\and
Landessternwarte, Universit\"at Heidelberg, K\"onigstuhl, D 69117 Heidelberg, Germany
% ...
}

   \date{Received: 4 September 2007 / Accepted: 25 October 2007}

 
  \abstract
   {}
   {The properties of the very high energy (VHE; E$>$100 GeV) 
$\gamma$-ray emission from the high-frequency
peaked BL\,Lac PG\,1553+113 are investigated.  An attempt is made
to measure the currently unknown redshift of this object.}
   {VHE Observations of PG\,1553+113 were made with the 
High Energy Stereoscopic System (HESS) in 2005 and 2006. 
H+K (1.45$-$2.45$\mu$m) spectroscopy of PG\,1553+113 
was performed in March 2006 with SINFONI, an integral field 
spectrometer of the ESO Very Large Telescope (VLT) in Chile.}
   {A VHE signal, $\sim$10 standard deviations, 
is detected by HESS during the 2 years of
observations (24.8 hours live time). 
The integral flux above 300 GeV is
$(4.6\pm0.6_{\rm stat}\pm0.9_{\rm syst}) \times 10^{-12}$ 
cm$^{-2}$\,s$^{-1}$,
corresponding to $\sim$3.4\% of the flux from the Crab Nebula
above the same threshold.  The time-averaged energy spectrum 
is measured from 225 GeV to $\sim$1.3 TeV, and is characterized
by a very soft power law (photon index of 
$\Gamma = 4.5\pm0.3_{\rm stat}\pm0.1_{\rm syst}$).
No evidence for any flux or spectral variations
is found on any sampled time scale within the VHE data.
The redshift of PG\,1553+113 could not be determined.
Indeed, even though the measured SINFONI spectrum 
is the most sensitive ever reported for this object at near infrared
wavelengths, and the sensitivity is comparable to the best 
spectroscopy at other wavelengths,
no absorption or emission lines were found 
in the H+K spectrum presented here.}
   {}


   \maketitle

\section{Introduction}

Evidence for VHE ($>$100 GeV) $\gamma$-ray emission from 
the active galactic nucleus (AGN) PG\,1553+113 was 
first reported by the HESS collaboration (\cite{HESS_discovery}) 
based on observations made in 2005. This evidence was later
confirmed (\cite{MAGIC_1553}) with MAGIC observations 
in 2005 and 2006. Similar to essentially all AGN 
detected at VHE energies, PG\,1553+113 is classified as 
a high-frequency peaked BL\,Lac \cite{classification}
and is therefore believed to possess the double-humped
broad-band spectral energy distribution (SED) typical of blazars.  
The low-energy (i.e. from the radio to the X-ray regime) portion of
the SED of PG\,1553+113 is well-studied,
including several simultaneous multi-wavelength observation 
campaigns (see, e.g., \cite{Example_1553_MWL}).
However, the only data in the high-energy hump are 
from HESS and MAGIC.  The measured VHE spectra are unusually soft 
(photon index $\Gamma$=4.0$\pm$0.6 and $\Gamma$=4.2$\pm$0.3, 
respectively) but the errors are large,
clearly requiring improved measurements before detailed 
interpretation of the complete SED is possible.

Further complicating any SED interpretation is the 
absorption of VHE photons (\cite{EBL_effect3,EBL_effect2})
by pair-production on the Extragalactic Background Light (EBL).
This absorption, which is energy dependent and increases strongly 
with redshift, distorts the VHE energy spectra observed from 
distant objects. For a given redshift and a given EBL model, the
effects of the latter on the observed spectrum can be 
reasonably accounted for during SED modeling.
Unfortunately, the redshift of PG\,1553+113 
is currently unknown\footnote{\cite{redshift_question} demonstrate the
catalog redshift of $z=0.36$ is incorrect.}.
To date no emission or absorption lines have been measured from
PG\,1553+113 despite more than ten observation campaigns 
with optical instruments, including EMMI at the NTT 
and FORS2 with the 8-meter VLT telescopes 
(\cite{Carangelo_03,no_lines}).  Lower limits 
of $z>0.09$ (\cite{no_lines}) and $z>0.3$ (\cite{Carangelo_03})
were determined from the lack of detected absorption/emission
lines, implying that the effect 
of the EBL is large in the observed VHE data. 
The absence of absorption and emission lines suggests 
that the non-thermal component of the emission 
from PG\,1553+113 is largely dominant 
over that of the host galaxy. 
This is consistent with the fact that, 
although some hints for a host galaxy have been suspected,
no clear detection has yet been found, even in 
Hubble Space Telescope (HST) images of 
PG\,1553+113 taken during the HST survey of 110 BL Lac 
objects \cite{Hubble_image}.  Interestingly,  approximately 
80\% (88/110) of the BL\,Lacs initially surveyed by HST have 
known redshifts, of which all 39 with $z<0.25$ and 
21 of the 28 with $0.25<z<0.6$ have their hosts resolved, suggesting
that the redshift of PG\,1553+113 is indeed large. 
Recently, using a re-analysis 
of the HST snapshot survey of BL\,Lacs,
\cite{z_extreme} claimed the 
dispersion of the absolute magnitude
of BL\,Lac hosts is sufficiently small that the 
measurement of the host-galaxy brightness
allows a reliable estimate of their redshift. 
With the assumption that there is no strong evolution,
these authors set a possible lower limit of $z>0.78$ for
PG\,1553+113. However, adding new STIS-HST data 
of $z>0.6$ BL\,Lacs to the snapshot survey, 
\cite{HST_new} found on the contrary that host
galaxies of BL\,Lacs evolve strongly, making 
any photometric redshift determination questionable.
A conservative upper limit ($z<0.74$) was 
determined (\cite{HESS_discovery}) from the photon spectrum 
measured by HESS.  The same limit was similarly determined
from the MAGIC spectral measurement (\cite{MAGIC_1553}).
Using both the MAGIC and HESS data,
a stronger upper limit of $z<0.42$ was later
reported \cite{Mazin_limit} based on the assumption 
that there is no break in the intrinsic VHE spectrum 
of the object.  The range of
allowed redshift is clearly large enough that 
the significant effects of EBL absorption cannot be reliably removed
from the observed VHE spectrum of PG\,1553+113. This correspondingly
makes modeling the high-energy portion of its SED unreliable.
A clear detection of the object's redshift would dramatically
improve the understanding of PG\,1553+113.  Further, if it
were found to be distant, its VHE spectrum could potentially 
provide strong constraints on the poorly-measured EBL
(assuming a reasonable intrinsic spectrum) 
and contribute to establishing 
the VHE $\gamma$-ray horizon.

Results from 17.2 hours of new HESS 
observations of PG\,1553+113 in 2006 
are reported here.  In addition, 
a re-analysis of the previously published
2005 HESS data (7.6 hours), with an improved calibration of 
the absolute energy scale of the detector, is presented.
HESS and the Suzaku X-ray satellite observed
the blazar simultaneously in July 2006.  
The HESS results from this epoch are
also discussed. This will enable, for the first time,
future modeling of an SED determined from
simultaneous observations at VHE and lower energies.  
Finally, results of a March 2006 VLT SINFONI spectroscopy
campaign to determine the redshift of PG\,1553+113
are also reported.

\section{HESS Observations and Analysis Technique}

A total of 30.3 hours of HESS observations 
were taken, typically in $\sim$28-minute data segments (runs),
on PG\,1553+113 in 2005 and 2006.
The pointing of the HESS telescopes was offset
$\pm$0.5$^{\circ}$ from the position of PG\,1553+113 
during the observations.
The useful exposure is 24.8 hours live time,
as 9 of the 66 runs fail
the HESS data-quality selection criteria.
The data are processed using the
standard HESS calibration \cite{calib_paper}
and analysis tools \cite{std_analysis}.
{\it Soft cuts} (\cite{HESS_discovery})\footnote{The {\it soft cuts}
were initially named the {\it spectrum cuts}  in Aharonian et al. (2006a).}
are applied to select candidate $\gamma$-ray events.
The {\it soft cuts} are selected instead of the
{\it standard cuts} since they result in a lower energy threshold.
Hence they are more appropriate given
the very soft VHE spectra previously measured 
by both HESS ($\Gamma=4.0$; \cite{HESS_discovery}) 
and MAGIC ($\Gamma=4.2$; \cite{MAGIC_1553}).
At the mean zenith angle of the observations, $37^{\circ}$,
the {\it soft cuts} result in an average post-analysis energy 
threshold\footnote{The threshold has
been corrected to account for the decreased 
optical efficiency of the HESS mirrors.} of 300 GeV.
Events falling within a circular region
of radius $\theta_{cut}$=0.14$^{\circ}$ centered
on PG\,1553+113 are considered on-source data.
The background (off-source data)
is estimated simultaneously to the on-source data from
events recorded within the same field of view (FoV) using
the {\it Reflected-Region} method \cite{bgmodel_paper}.
Equation (17) from \cite{lima} is used to calculate the
significance of any excess.  The energy of each event 
is corrected \cite{HESS_crab} using efficiencies determined 
from simulated and observed muons.  This correction accounts for
long-term variations in the absolute energy scale
due to a changing optical throughput of the HESS system.
The HESS results have been verified using independent calibration 
and analysis chains.

\section{HESS Results}

  \begin{table*}
      \begin{minipage}[t]{2.0\columnwidth}
      \caption{Results of the HESS observations of PG\,1553+113.}
         \label{results}
        \centering
	\renewcommand{\footnoterule}{}
         \begin{tabular}{c c c c c c c c c c c c c}
            \hline\hline
            \noalign{\smallskip}
            Dark & MJD & MJD & Time & On & Off & $\alpha$ & Excess & Sig & I($>$300 GeV)\footnote{The systematic error on the observed integral flux above 300 GeV is 20\% and is not shown.} & Crab\footnote{The integral flux percentage is
calculated relative to the Crab Nebula flux above 300 GeV.} & $\chi^2$\,,\,NDF\footnote{The $\chi^2$, degrees of 
	freedom (NDF), and $\chi^2$ probability P($\chi^2$) are given
        for a fit of a constant to I($>$300 GeV) binned
        nightly within a dark period, or monthly within a year,
	or yearly within the total.} & 
P($\chi^2$)\\
            Period & First & Last & [h] & & & & &  [$\sigma$] & [10$^{-12}$\,cm$^{-2}$\,s$^{-1}$] & \% & & \\
            \noalign{\smallskip}
            \hline
            \noalign{\smallskip}
            04/2005 & 53492  & 53507 & 4.9  & 1210 & 8154 & 0.125 & 191 & 5.5 & $4.78\pm1.18$  & 3.5 & 2.7\,,\,4 & 0.61\\
            08/2005 & 53609 & 53614 & 2.7  & 491 & 3462  & 0.125 & 58  & 2.5 & $10.19\pm5.46$ & 7.5 & 4.9\,,\,2 & 0.09\\
            04/2006 & 53849 & 53860 & 7.1  & 1811 & 12742 & 0.125 & 218 & 5.0 & $4.08\pm1.10$  & 3.0 & 5.3\,,\,8 & 0.72\\
            07/2006 & 53938 & 53943  & 10.1 & 2236 & 15341 & 0.125 & 318 & 6.7 & $4.33\pm0.94$  & 3.2 & 3.1\,,\,5 & 0.68\\
           \noalign{\smallskip}
            \hline
            \noalign{\smallskip}
            2005 & 53492 & 53614 & 7.6  & 1701 & 11616 & 0.125 & 249 & 6.0 & $5.44\pm1.23$  & 4.0 & 0.9\,,\,1  & 0.33\\
            2006 & 53849 & 53943 & 17.2 & 4047 & 28083 & 0.125 & 536 & 8.3 & $4.22\pm0.72$  & 3.1 & 0.03\,,\,1 & 0.86\\
            \noalign{\smallskip}
            \hline
            \noalign{\smallskip}
            Total & 53492 & 53943 & 24.8 & 5748 & 39699 & 0.125 & 785 & 10.2 & $4.56\pm0.62$  & 3.4 & 0.7\,,\,1 & 0.39\\             
            \noalign{\smallskip}
            \hline
       \end{tabular}
     \end{minipage}
   \end{table*}

   \begin{figure}
   \centering
      \includegraphics[width=0.45\textwidth]{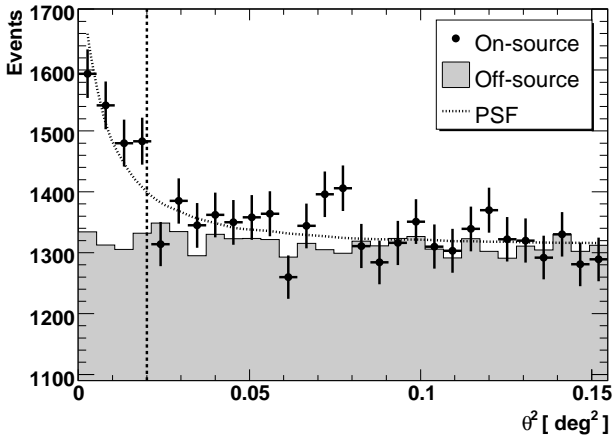}
      \caption{Distribution of $\theta^2$ for on-source 
        events (points) and
        normalized off-source events (shaded) from observations
        of PG\,1553+113.  The dashed curve represents
	the $\theta^2$ distribution expected for
	a point source of VHE $\gamma$-rays at $40^{\circ}$
	zenith angle with a photon index $\Gamma=4.46$.
	The vertical line represents the cut 
        on $\theta^2$ applied to the data.}
         \label{thtsq_plot}
   \end{figure}

The results of the HESS observations for each dark period in 2005 and 2006,
as well as relevant totals for the observations, are given in Table~\ref{results}.
The table shows the dark periods in which PG\,1553+113 was observed,
the MJD of the first and last night of observations, 
the dead time corrected observation time, the number of on- and off-source events, 
the on/off normalization ($\alpha$), the measured excess, and 
the significance of the excess.
A significant excess of events from the direction of
PG\,1553+113 is clearly detected in each year of
HESS data taking.  A total of 785 excess events, corresponding
to a statistical significance of 10.2 standard deviations ($\sigma$),
is detected in the complete data set.
Figure~\ref{thtsq_plot} shows the on-source and normalized off-source
distributions of the square of the angular difference between
the reconstructed shower position and
the source position ($\theta^{2}$) for all observations. 
There is a clear point-like
excess of on-source events at small values of $\theta^{2}$, 
corresponding to the observed signal,
and the background is approximately 
flat in $\theta^{2}$ as expected.
The peak of a two-dimensional Gaussian fit to a sky map 
of the observed excess is located at 
$\alpha_{\rm J2000}=15^{\mathrm h}55^{\mathrm m}44.7^{\mathrm s}\pm3.0^{\mathrm s}_{\rm stat}\pm1.3^{\mathrm s}_{\rm syst}$,
$\delta_{\rm J2000}=11^{\circ}11'41''\pm53''_{\rm stat}\pm20''_{\rm syst}$.
The excess, HESS\,J1555+111, is coincident with the position 
($\alpha_{\rm J2000}=15^{\mathrm h}55^{\mathrm m}43.0^{\mathrm s}$, 
$\delta_{\rm J2000}=11^{\circ}11'24.4''$) of PG\,1553+113 \cite{position_1553}.
The upper limit (99\% confidence level) on the extension
of HESS\,J1555+111 is $1.1'$.

\subsection{Spectral Studies}

\begin{figure}
   \centering
      \includegraphics[width=0.45\textwidth]{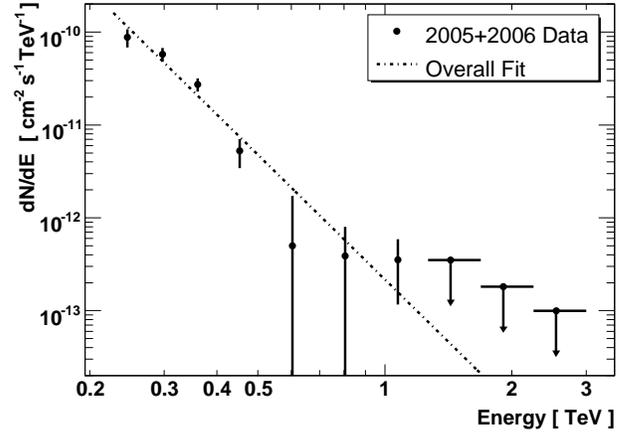}
      \caption{Overall VHE energy spectrum observed from PG\,1553+113. 
	The dashed line represents the best $\chi^2$ fit of a power law to
        the observed data.  The upper limits are at the 99\% confidence
	level \cite{UL_tech}. Only the statistical errors are shown.}
         \label{spectrum_plot}
   \end{figure}

The photon spectrum for the entire data set is shown 
in Figure~\ref{spectrum_plot}. These data can be fit
($\chi^2$ of 8.4 for 5 degrees of freedom) by
a power-law function d$N$/d$E$ = 
$I_{\circ} \hspace{0.1cm} (E \hspace{0.1cm}  / \hspace{0.1cm} {\rm 300 \hspace{0.1cm} GeV})^{-\Gamma}$)
with a photon index  
$\Gamma=4.46\pm0.34_{\rm stat}\pm0.10_{\rm syst}$.
Fits of either a power law with an exponential cut-off
or a broken power law do not yield significantly
better $\chi^2$ values.  It should be noted that each
of the three highest energy points, $E>0.6$ TeV, in 
Figure~\ref{spectrum_plot} have statistical significance less
than 2$\sigma$.  However, removing these points from the fit
does not change $\Gamma$ significantly.

   \begin{figure}
   \centering
      \includegraphics[width=0.45\textwidth]{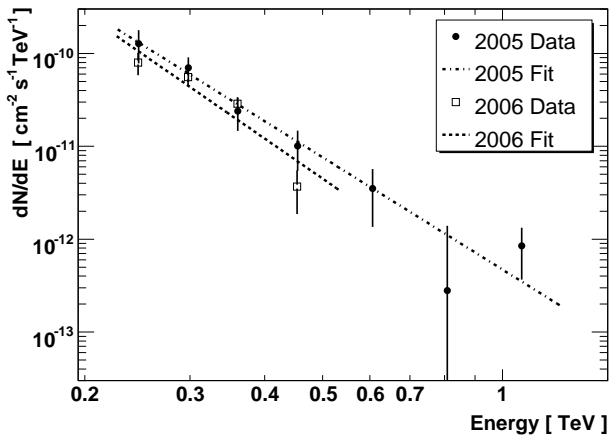}
      \caption{Annual VHE energy spectra observed by HESS 
	from PG\,1553+113. 
	Each line represents the best $\chi^2$ fit of a power law to
        the observed data. Only the statistical errors are shown.}
         \label{annual_spectra}
   \end{figure}

   \begin{table*}
      \begin{minipage}[t]{2.0\columnwidth}
      \caption{Results of the best $\chi^2$ fit to the
	annual and overall spectra of PG\,1553+113 measured by HESS.}
         \label{annual_results}
        \centering
	\renewcommand{\footnoterule}{}
         \begin{tabular}{c c c c c c c c}
            \hline\hline
            \noalign{\smallskip}
             Epoch & E$_{\rm min}$ & E$_{\max}$ & $\Gamma$ & $I_{\circ}$ & $\chi^2$ & NDF & P($\chi^2$)\\
		& [TeV] & [TeV] & & [$10^{-11}$ cm$^{-2}$\,s$^{-1}$\,TeV$^{-1}$] & & & \\
            \noalign{\smallskip}
            \hline
            \noalign{\smallskip}
	    2005 (AH06)\footnote{The AH06 entry is the
	previously published HESS result (\cite{HESS_discovery})
	for 2005. The AH06 entry was not corrected for
	long-term changes in the optical efficiency of the system.}
& 0.185 & 0.585 & $3.98\pm0.63_{\rm stat}\pm0.10_{\rm syst}$ & $2.59\pm0.47_{\rm stat}\pm0.52_{\rm syst}$ & 1.7 & 2 & 0.42 \\
            \noalign{\smallskip}
            \hline
            \noalign{\smallskip}
	    2005\footnote{The 2005 entry corresponds to exactly the same data as presented in AH06,
	but with a correction (see text) applied to account for
	optical efficiency changes within the data. The 2006 and
	total entries also have this correction applied.} 
		 & 0.225 & 1.265 & $4.01\pm0.60_{\rm stat}\pm0.10_{\rm syst}$ & $5.92\pm1.19_{\rm stat}\pm1.18_{\rm syst}$ & 2.1 & 5 & 0.84\\
	    2006 & 0.225 & 0.534 & $4.45\pm0.48_{\rm stat}\pm0.10_{\rm syst}$ & $4.35\pm0.54_{\rm stat}\pm0.87_{\rm syst}$ & 8.5 & 2 &  0.014\\
            \noalign{\smallskip}
            \hline
            \noalign{\smallskip}
	    Total & 0.225 & 1.265 & $4.46\pm0.34_{\rm stat}\pm0.10_{\rm syst}$ & $4.68\pm0.49_{\rm stat}\pm0.94_{\rm syst}$ & 8.4 & 5 & 0.13\\
          \noalign{\smallskip}
            \hline
       \end{tabular}
     \end{minipage}
   \end{table*}

The significant photon statistics from each year of observations
allows the determination of annual spectra.  These spectra are shown
in Figure~\ref{annual_spectra}.  The results of fits to the data
of a power law for both years, as well as for the total,
are shown in Table~\ref{annual_results}.  The
lower and upper energy bounds, photon index ($\Gamma$), differential flux 
normalization at 300 GeV ($I_{\circ}$), $\chi^2$, 
degrees of freedom (NDF), and $\chi^2$ probability P($\chi^2$)
for each fit are given in the table. There are no
significant changes in the spectral shape during the HESS observations.
By including one negative point at $\sim$600 GeV, and two more positive
points at higher energies, the 2006 spectrum can be fit over the 
same range as either the 2005 or the overall spectrum.  This 
fit yields $\Gamma \approx 4.9$, explaining why the overall photon 
index is slightly softer than the 2005 and 2006 
indices reported in Table~\ref{annual_results}.

\subsection{Integral Flux Studies}

\begin{figure}
   \centering
      \includegraphics[width=0.45\textwidth]{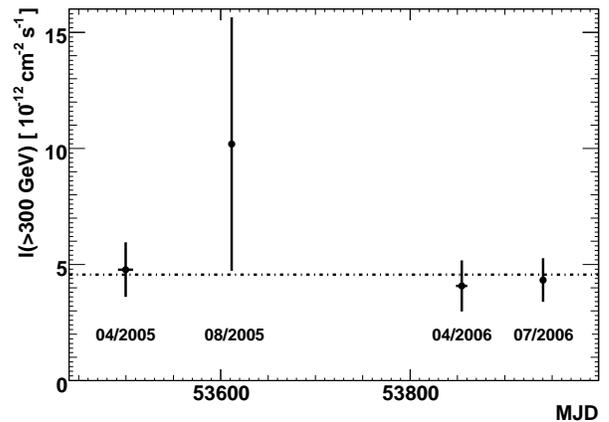}
      \caption{Integral flux, I($>$300 GeV), measured by HESS
from PG\,1553+113 during each dark period of observations.
The horizontal line represents the average flux for all 
the HESS observations. For each point the time-averaged $\Gamma=4.46$ 
is assumed. Only the statistical errors are shown.}
         \label{monthly_plot}
   \end{figure}

\begin{figure*}
   \centering

  $\begin{array}{c@{\hspace{0.5cm}}c}

  \multicolumn{1}{l}{\mbox{\bf }} &
        \multicolumn{1}{l}{\mbox{\bf }} \\ [0cm]

  \includegraphics[width=0.40\textwidth]{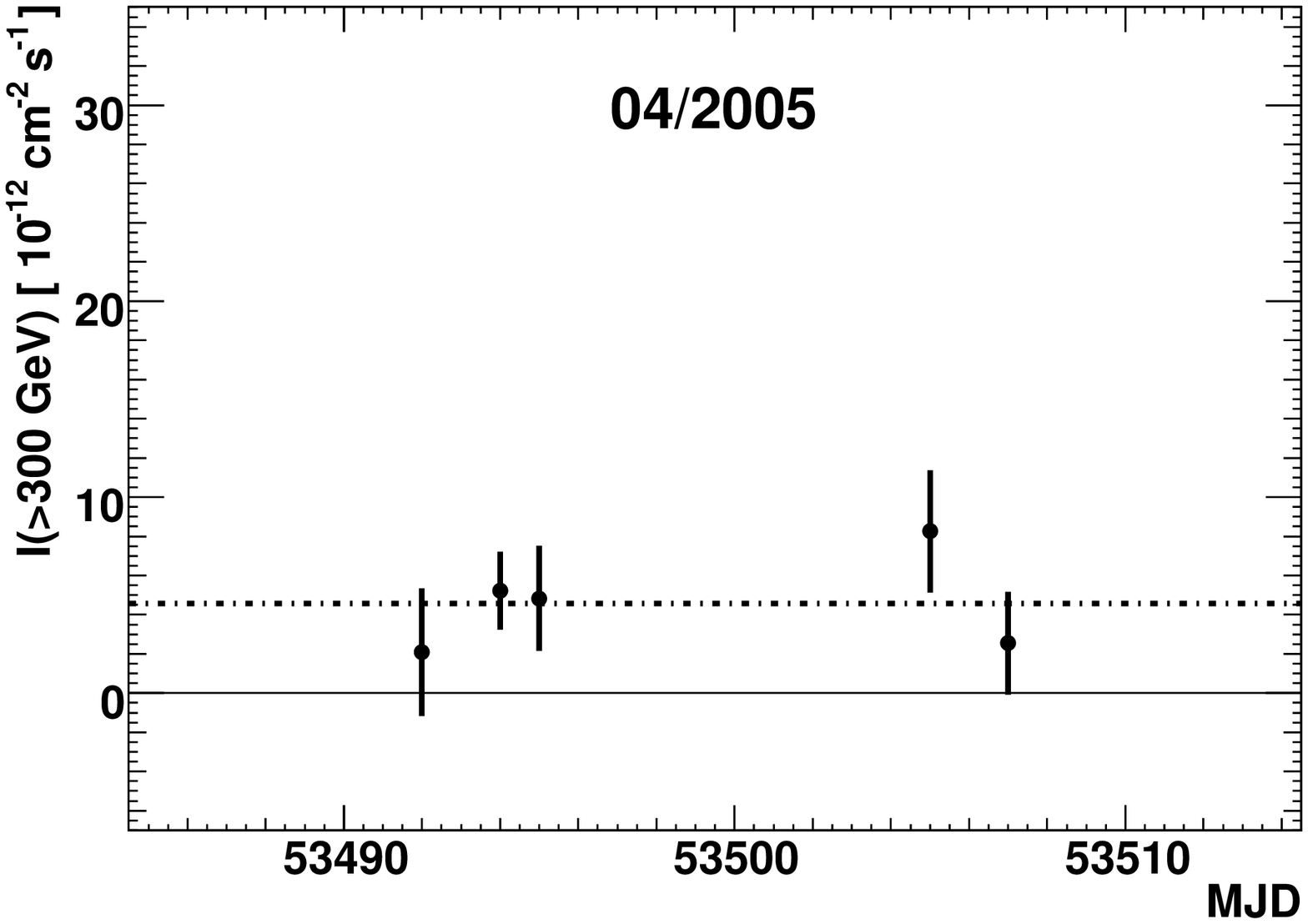} &
  \includegraphics[width=0.40\textwidth]{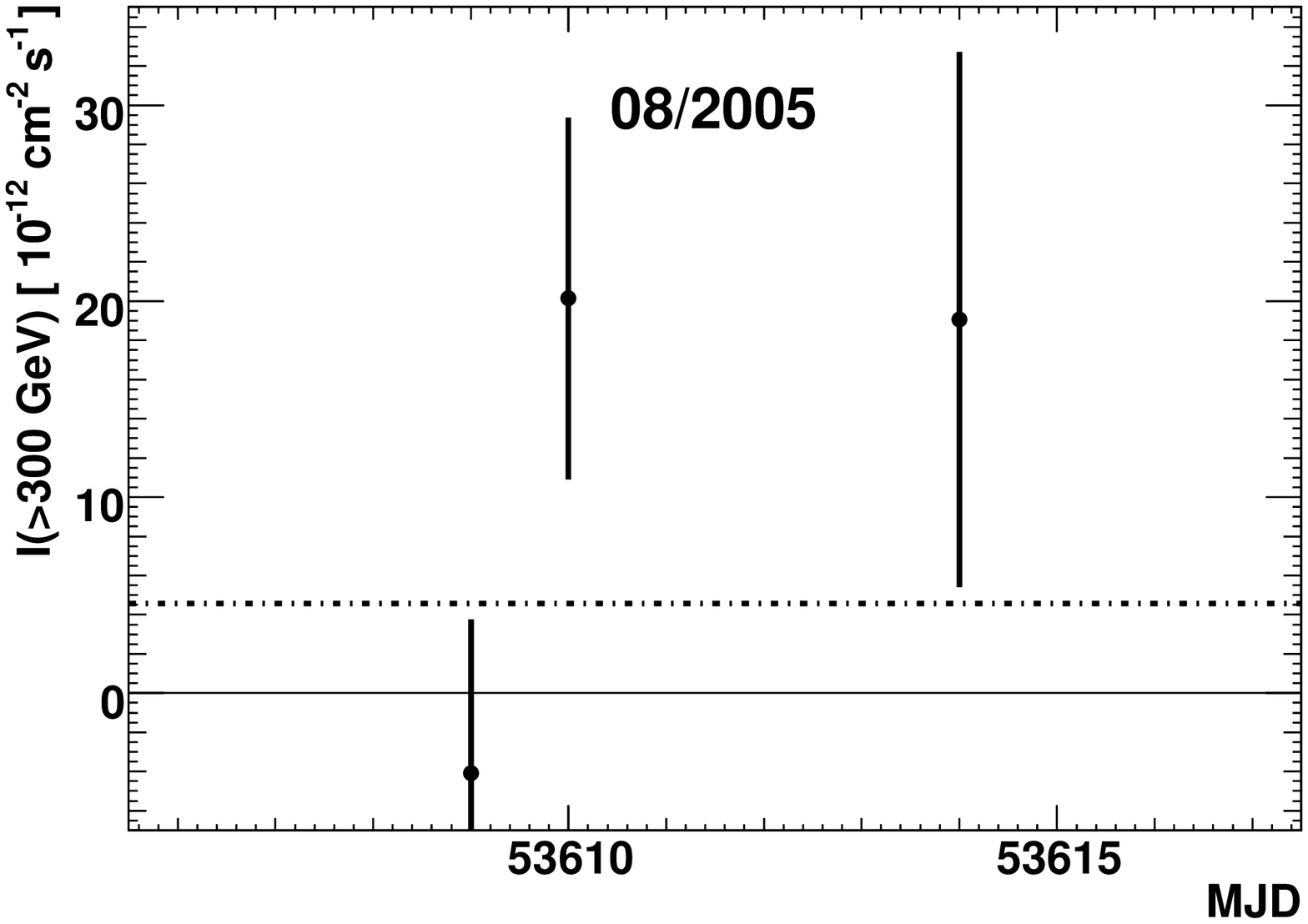} \\ [0.0cm]
  \mbox{\bf (a)} & \mbox{\bf (b)} \\  [0.2cm]
  \includegraphics[width=0.40\textwidth]{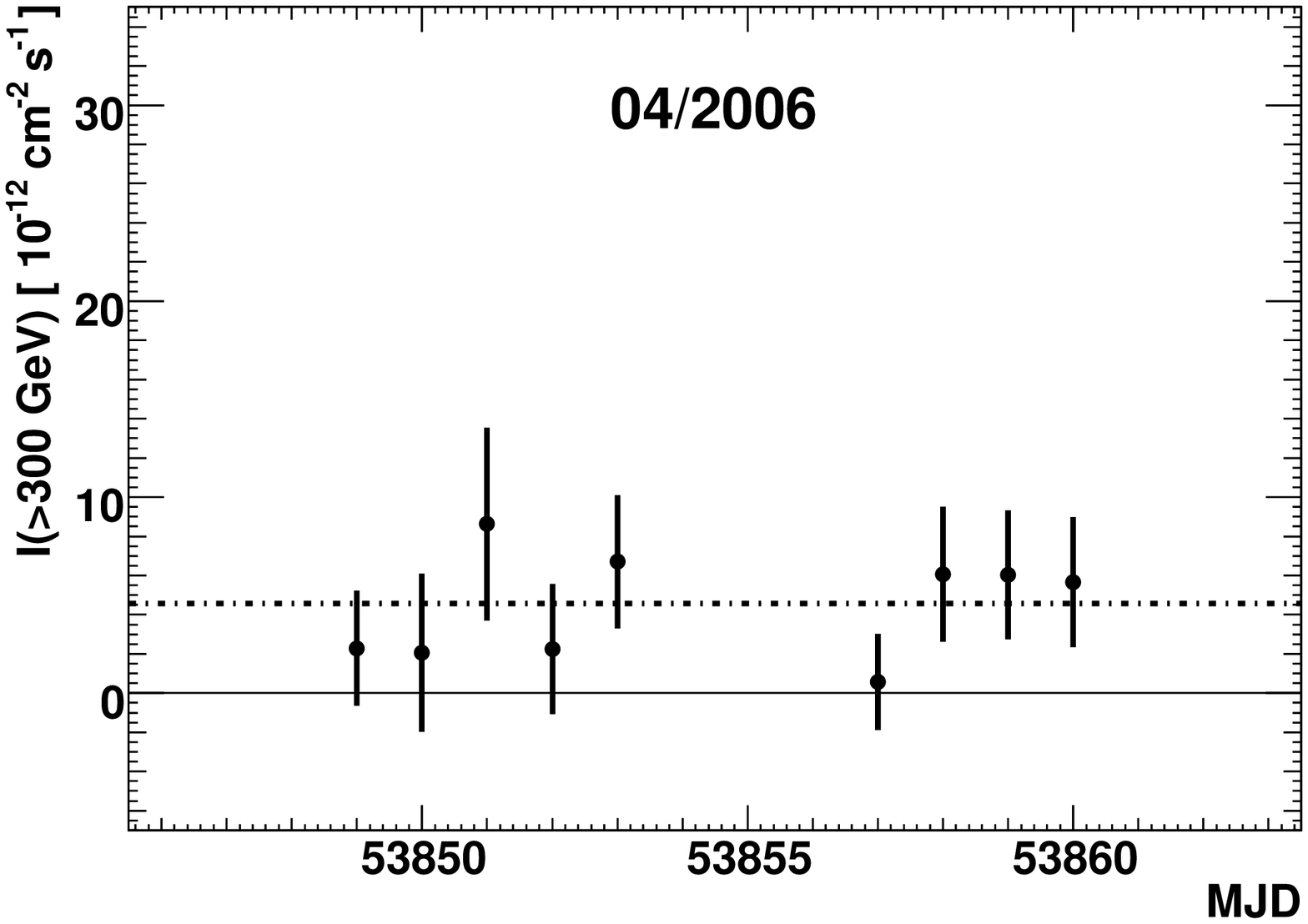} &
  \includegraphics[width=0.40\textwidth]{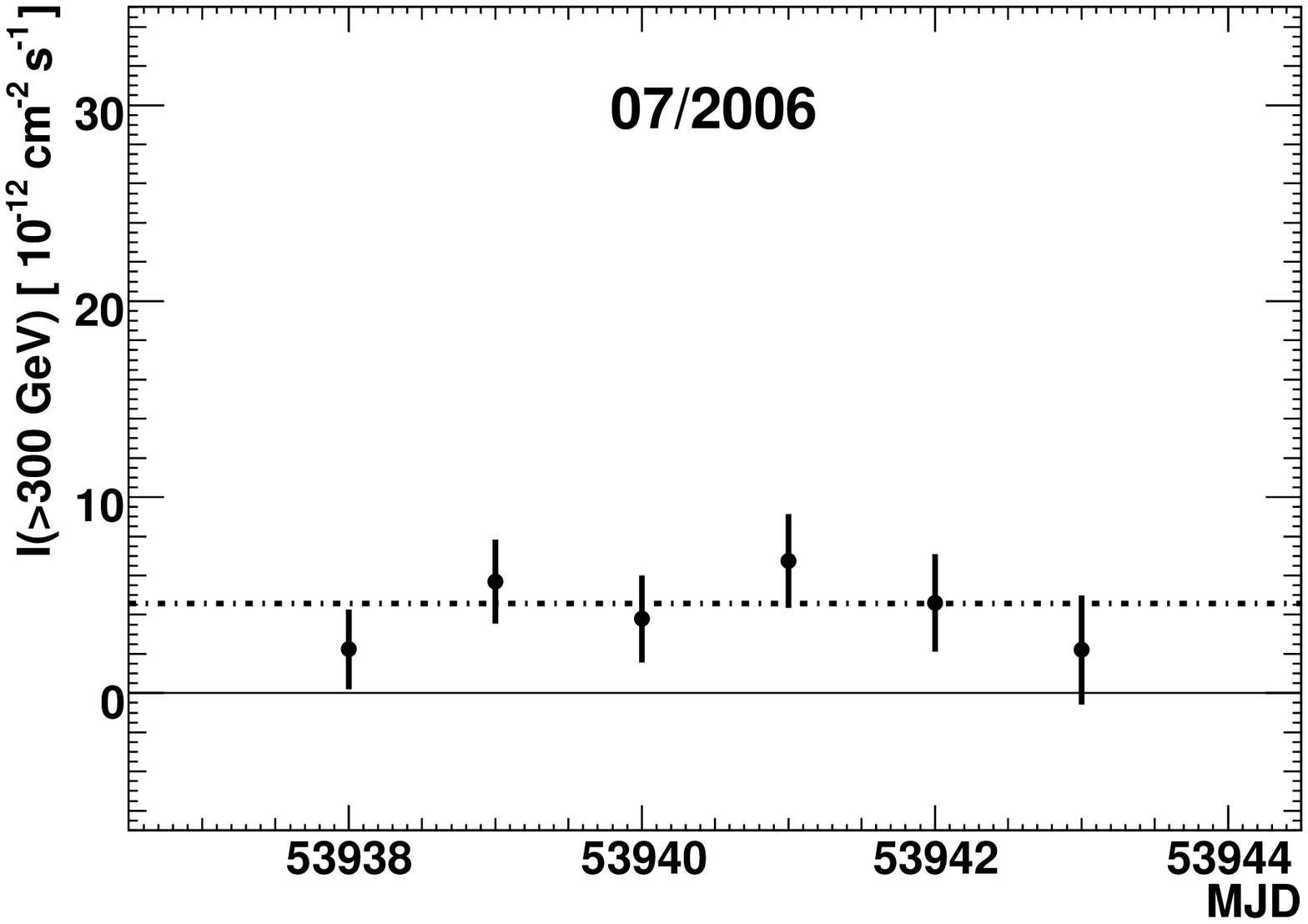} \\ [0.0cm]
  \mbox{\bf (c)} & \mbox{\bf (d)}  \\ [0.2cm]

  \end{array}$
      \caption{Integral flux, I($>$300 GeV), measured by HESS
from PG\,1553+113 during each night of observations.  The plots a, b, c, and d,
represent the April 2005, August 2005, April 2006 and July 2006 dark
periods, respectively.  The horizontal line represents the 
average flux for all the HESS observations.
For each point the time-average $\Gamma=4.46$ is assumed.
Only the statistical errors are shown.}
         \label{nightly_plots}
   \end{figure*}

All HESS fluxes (i.e. annual, dark period and nightly 
values) throughout this article are
calculated assuming the measured time-average photon index of
$\Gamma=4.46$.  Assuming a different value (i.e. $4.0 < \Gamma < 5.0$)
has less than a 5\% effect on the flux.
The observed integral flux above 300 GeV for the entire data set is
I($>$300 GeV) = $(4.56\pm0.62_{\rm stat}\pm0.91_{\rm syst}) \times 10^{-12}$ 
cm$^{-2}$\,s$^{-1}$.  This corresponds to $\sim$3.4\% of I($>$300 GeV)
determined from the HESS Crab Nebula spectrum\cite{HESS_crab}.
The integral flux, I($>$300 GeV), is shown in Table~\ref{results}
for each year of observations, as well as for each dark period.
The $\chi^2$ and corresponding probability 
shown in the table is for a fit of a constant to
the data when binned by nights within each dark period, by dark periods within
a year, and by year within the total observations.  There
are no indications for flux variability on any time scale within the HESS
data.  Figures \ref{monthly_plot} and \ref{nightly_plots} 
show the flux measured for each dark period and night, respectively.

\subsection{Effect of the Optical Efficiency Correction}

\begin{figure}
   \centering
      \includegraphics[width=0.45\textwidth]{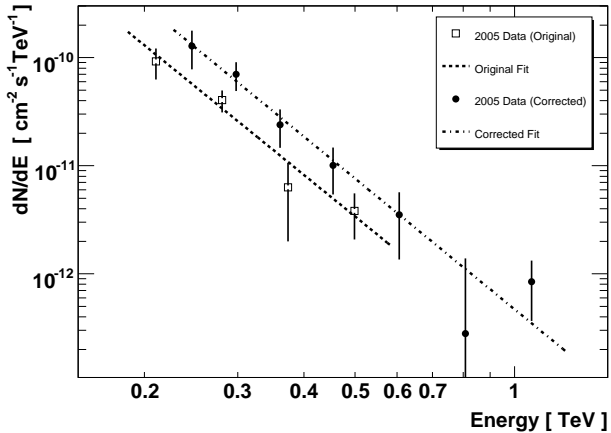}
      \caption{Spectrum measured by HESS from PG\,1553+113 in
	2005 compared to the previously published version 
	(\cite{HESS_discovery}). The integral
	flux above 300 GeV is three times higher. Only the statistical errors
	are shown.}
         \label{muoncorr_plot}
   \end{figure}

The data previously published (\cite{HESS_discovery})
for HESS observations  of PG\,1553+113 in 2005 were
not corrected for long-term changes in the optical sensitivity
of the instrument.  Relative to a virgin telescope, the total
optical throughput was decreased by 29\% in 2005 and 33\%
in 2006.  These losses are due to the reduced reflectivity of both
the mirrors and Winston cones, and to the accumulation of dust
on the optical elements.
Comparing (see Table~\ref{annual_results})
the spectrum for 2005 determined here to that 
which was previously published for the same data shows
significant differences in the flux normalization 
($I_{\circ}$) but not in the photon index ($\Gamma$). 
Figure ~\ref{muoncorr_plot} illustrates
this difference which is a result of correcting the energy
of individual events for the relative 
optical efficiency of the system, determined from simulated and
observed muon events, as described earlier.  
The flux measured in 2005 is therefore three times
higher than previously published\footnote{The flux was reported
above a threshold of 200 GeV.  Extrapolating the earlier result, 
using $\Gamma=4.0$ as reported, yields 
I($>$300 GeV) = $(1.78\pm0.45_{\rm stat}\pm0.35_{\rm syst}) \times 10^{-12}$ 
cm$^{-2}$\,s$^{-1}$.}. The large difference is due to the 
steep spectrum of the source. For harder spectrum sources
the effect is much smaller.  

\subsection{Comparison to MAGIC Results}

Table~\ref{result_compare} shows the photon index and integral flux
for PG\,1553+113 measured by both the MAGIC (\cite{MAGIC_1553}) 
and HESS collaborations.  As the data were not simultaneously measured, 
one might not expect the measured quantities 
to agree due to variability arguments.  In addition, the spectra from
the two instruments are measured over different energy ranges
which could result in different spectral slopes if the spectrum
were curved. Nevertheless, the spectral slope ($\Gamma$) 
measured by both instruments agree within errors in all epochs.  
Further, the flux above 200 GeV 
measured by both HESS and MAGIC
in 2005 is consistent\footnote{Previously the 2005 HESS result
was $\sim$4 times lower than the 2005 MAGIC result.  The agreement
is solely the result of the optical efficiency correction.}.  
However the flux observed in 2006 by MAGIC is lower 
than the HESS value.  It should be
noted that in 2006, all the HESS data were taken after the final
published MAGIC observations, whereas a good fraction of the 2005
data from both instruments are quasi-simultaneous.

  \begin{table*}
      \begin{minipage}[t]{2.0\columnwidth}
      \caption{Photon indices and integral fluxes measured by HESS and MAGIC.}
         \label{result_compare}
        \centering
	\renewcommand{\footnoterule}{}
         \begin{tabular}{c c c c c}
            \hline\hline
            \noalign{\smallskip}
		& HESS\footnote{The systematic error on the HESS $\Gamma$ is 0.1.} 
& HESS\footnote{The HESS fluxes are extrapolated from the values in Table~\ref{results}
to above 200 GeV using the time-average $\Gamma=4.46$. The systematic error on the HESS flux is 20\%.} 
& MAGIC & MAGIC\\
             Epoch & $\Gamma$  & I($>$200 GeV)
&  $\Gamma$ & I($>$200 GeV)\\
		   & & [$10^{-11}$ cm$^{-2}$\,s$^{-1}$] & & [$10^{-11}$ cm$^{-2}$\,s$^{-1}$]\\
            \noalign{\smallskip}
            \hline
            \noalign{\smallskip}
	    2005 & $4.01\pm0.60_{\rm stat}$ & $2.21\pm0.50_{\rm stat}$ & $4.31\pm0.45_{\rm stat}$ & $2.0\pm0.6_{\rm stat}$\\
	    2006 & $4.45\pm0.48_{\rm stat}$ & $1.72\pm0.29_{\rm stat}$ & $3.95\pm0.35_{\rm stat}$ & $0.6\pm0.2_{\rm stat}$\\
            \noalign{\smallskip}
            \hline
            \noalign{\smallskip}
	    2005$+$2006 & $4.46\pm0.34_{\rm stat}$ & $1.85\pm0.25_{\rm stat}$ & $4.21\pm0.25_{\rm stat}$ & $1.0\pm0.4_{\rm stat}$\\
          \noalign{\smallskip}
            \hline
       \end{tabular}
     \end{minipage}
   \end{table*}

\subsection{HESS Results during Suzaku Observations}

The Suzaku X-ray satellite \cite{Suzaku_info} observed PG\,1553+113 
from July 24, 2006, 14:26 UTC to July 25, 2006, 19:17 UTC with an
average efficiency of $\sim$53\%
(Suzaku Observation Log: http://www.astro.isas.ac.jp/suzaku/index.html.en).
HESS observed PG\,1553+113 for 3.1 hours live time on the two dates
of Suzaku observations. The HESS observation log on these dates,
henceforth called the Suzaku epoch, is shown in 
Table~\ref{HESS_suzaku_log}. The log contains the start and stop times of the seven
HESS observation runs from the two dates of Suzaku observations.  In addition,
the live time, observed excess and corresponding significance 
for each run are shown. All runs, except the last, are simultaneous with
Suzaku observations.  The last HESS run begins two minutes after the end
of the Suzaku pointing.

  \begin{figure*}
   \centering
  $\begin{array}{c@{\hspace{0.5cm}}c}

  \multicolumn{1}{l}{\mbox{\bf }} &
        \multicolumn{1}{l}{\mbox{\bf }} \\ [0cm]

  \includegraphics[width=0.45\textwidth]{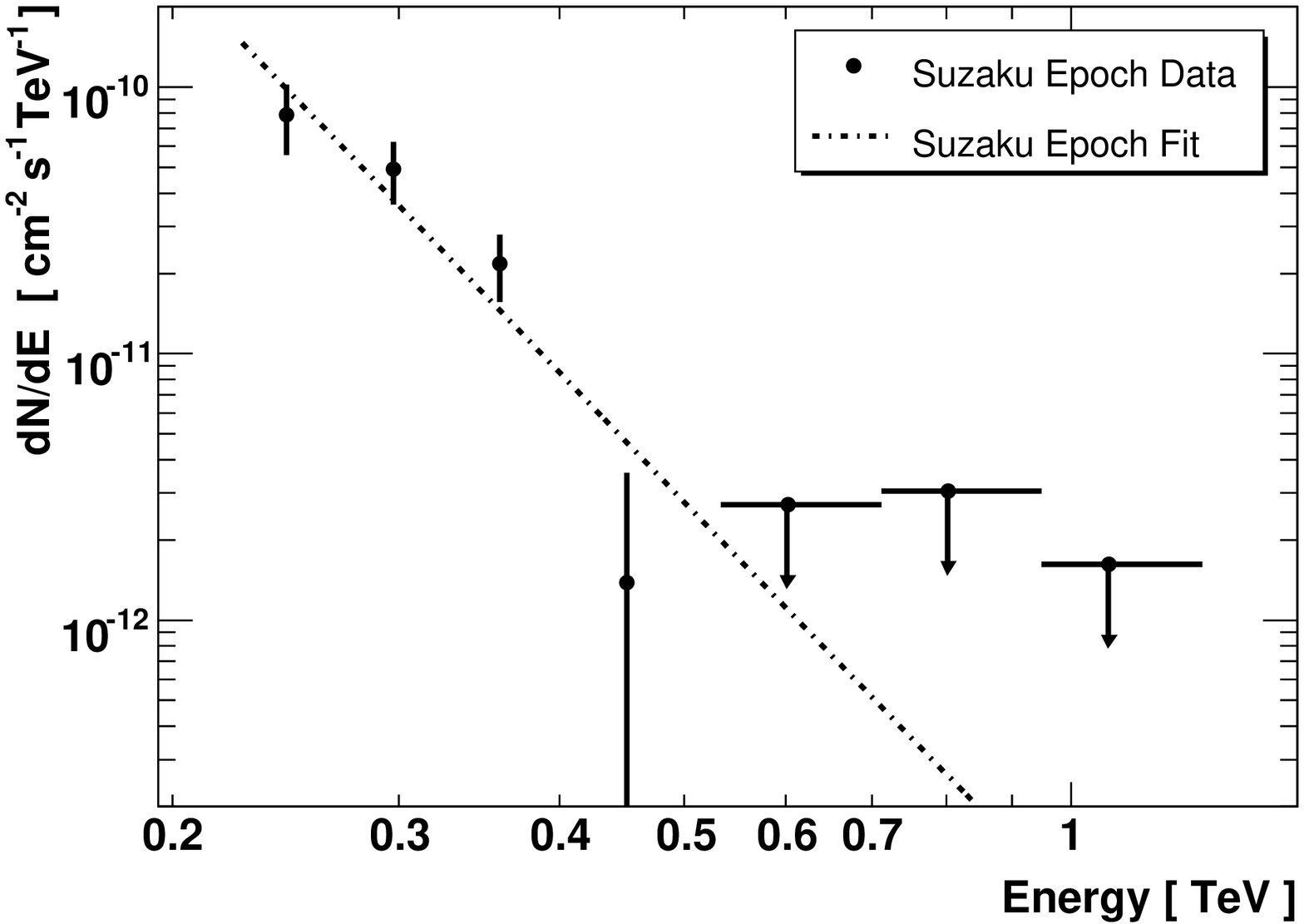} &
  \includegraphics[width=0.45\textwidth]{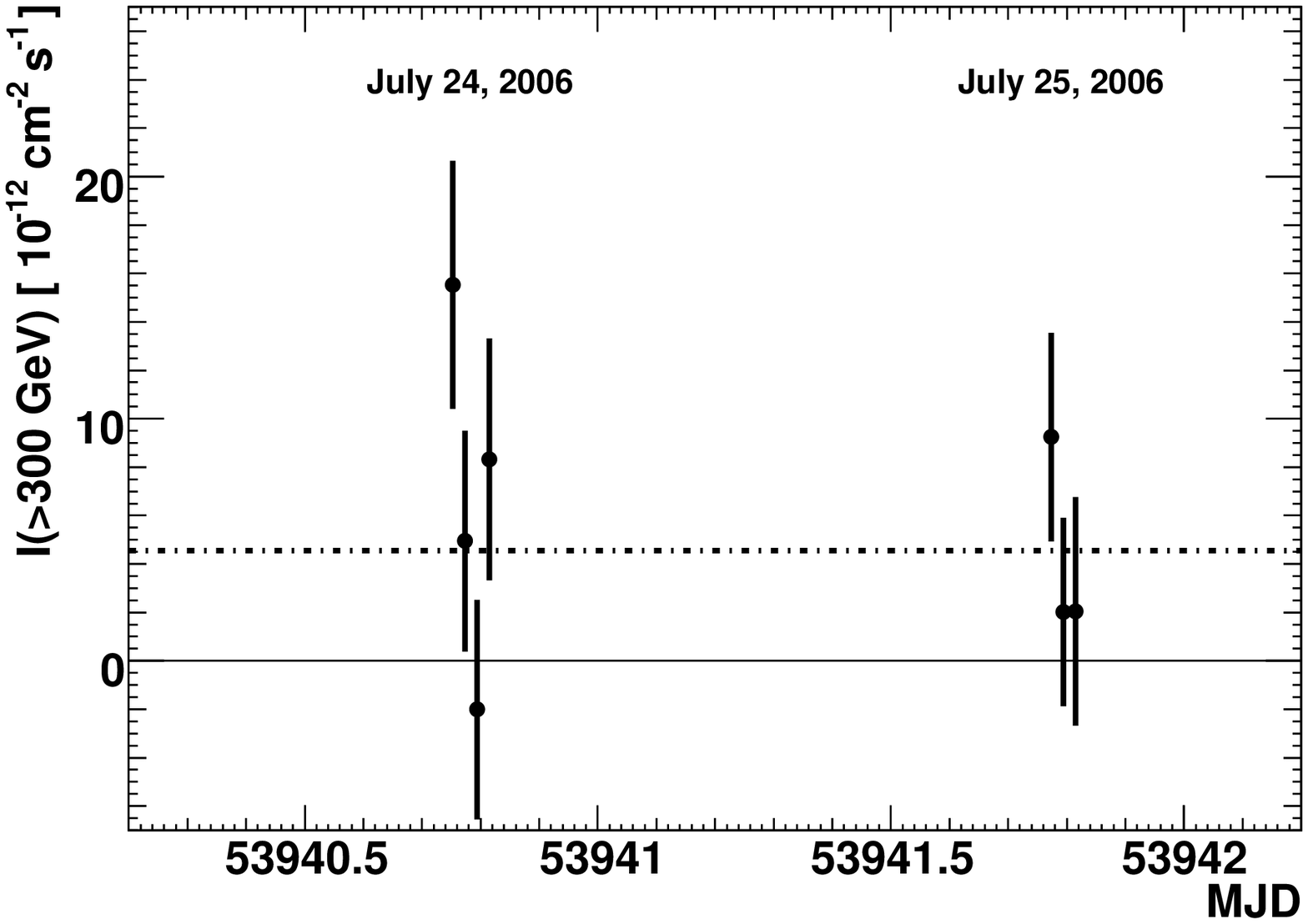} \\ [0.0cm]
  \mbox{\bf (a)} & \mbox{\bf (b)} \\  [0.2cm]
  \end{array}$

\caption{(a) Spectrum observed by HESS from PG\,1553+113
during observations in July 2006. 
The dashed line represents the best $\chi^2$ fit of a power law to
the observed data.  The upper limits are at the 99\% confidence
level \cite{UL_tech}.
(b) Integral flux, I($>$300 GeV), measured by HESS
from PG\,1553+113 in each run taken during simultaneous
Suzaku observations.  
The horizontal line represents the average flux for all 
the HESS observations. For each point in the light curve
the time-average $\Gamma=4.46$ is assumed. 
For both plots, only the statistical errors are shown.} 
         \label{MWL_info}
   \end{figure*}

During the Suzaku epoch, a total of 659 on-source events 
and 4462 off-source events are measured 
with an on-off normalization of 0.125,
corresponding to an excess of 101 events (3.9$\sigma$). 
Due to the low statistics it is not possible to 
produce a $\gamma$-ray spectrum from the data.
To provide a quasi-simultaneous measurement
for future modeling, the $\gamma$-ray spectrum observed
during the July 2006 dark period is shown in Figure~\ref{MWL_info}a. 
The best $\chi^2$ fit of a power law to the July 2006 data yields
$\Gamma = 5.0\pm0.7$, $I_{\circ} = 3.6\pm0.6 \times 10^{-13}$
cm$^{-2}$\,s$^{-1}$\,TeV$^{-1}$, and a $\chi^2$ of 4.8 for
2 degrees of freedom.  Only the statistical errors are
presented for $\Gamma$ and $I_{\circ}$.  The 
systematic errors are 0.1 and 20\%, respectively.

The average flux\footnote{Assuming the poorly-measured 
$\Gamma$=5.01 from July 2006, 
instead of the time-averaged $\Gamma=4.46$,
increases the flux by only $\sim$3.5\%.} 
during the Suzaku epoch is I($>$300 GeV) = 
$(5.8\pm1.7_{\rm stat}\pm1.2_{\rm syst}) \times 10^{-12}$ 
cm$^{-2}$\,s$^{-1}$.  The flux for each of the nights in the 
Suzaku epoch can be seen in Figure~\ref{nightly_plots}d.   
In addition, the flux for each HESS run ($\sim$28 min)
during the two nights is shown in Figure~\ref{MWL_info}b.  There are
no significant variations of the run-wise flux during each night,
or of the nightly flux during the Suzaku epoch.

 \begin{table}
      \begin{minipage}[t]{\columnwidth}
      \caption{HESS observation log during the Suzkau epoch.}
         \label{HESS_suzaku_log}
        \centering
	\renewcommand{\footnoterule}{}
         \begin{tabular}{c c c c c}
            \hline\hline
            \noalign{\smallskip}
	Start & Stop & Time & Excess & Significance\\
	${\rm [MJD]}$ & [MJD] & [h] & & [$\sigma$]\\
            \noalign{\smallskip}
            \hline
            \noalign{\smallskip}
53940.74231 & 53940.76185 & 0.453 & 27   & 2.5\\
53940.76328 & 53940.78277 & 0.436 & 20   & 1.9\\
53940.78431 & 53940.80385 & 0.437 & 7    & 0.7\\
53940.80531 & 53940.82480 & 0.435 & $-$2 & $-$0.2\\
\\
53941.76336 & 53941.78291 & 0.444 & 27   & 3.1\\
53941.78439 & 53941.80392 & 0.444 & 14   & 1.6\\
53941.80542 & 53941.82497 & 0.443 & 7    & 0.8\\
            \noalign{\smallskip}
            \hline
       \end{tabular}
     \end{minipage}
   \end{table}

\section{SINFONI Near-IR Spectroscopy}

The determination of the redshift of an AGN is
generally based upon the detection of emission or
absorption lines in its spectrum. In the unified model 
of AGN \cite{AGN_model} emission lines are 
generated by ionization from the central source.  
However in BL\,Lac objects like
PG\,1553+113 the line-equivalent
width of these emission lines is reduced, possibly even below detectable
levels, by the strong beaming of the jet continuum, that is
caused by the alignment of the 
relativistic jet with the line of sight to Earth. Absorption lines 
can be produced either from spectral features of the 
stellar population of the host galaxy or
from intervening halos as in the case of quasars.  
The detectability of stellar spectral features 
depends inversely on the brightness of the central source.
Therefore, the probability of detection of any spectral features 
is likely optimal in the near infrared (IR) wavelength range. Here,
the stellar contribution to the overall continuum emission is much 
larger than at optical or mid-IR wavelengths 
due to the shape of the SEDs of the AGN and the stars.
In addition, the extinction affecting the stellar light is much
smaller at near-IR wavelengths than at
optical (e.g. A$_H$ $\sim$ 0.17 A$_V$) wavelengths \cite{schlegel}.

In an attempt to detect absorption features from the host galaxy or
emission lines from the AGN,  H+K (1.50--2.40$\mu$m) spectroscopy 
of PG\,1553+113 was performed with SINFONI, an integral field 
spectrometer mounted at Yepun, Unit Telescope 4 of the 
ESO Very Large Telescope in Chile. The observations were 
performed in seeing-limited mode using the 125~mas
pixel scale, with a spectral resolution of $R_{H+K}\sim 1500$. The
seeing was typically 0\farcs7 FWHM during the 2.5 total
hours of on-source observations performed on 
March 9, 2006 and March 15, 2006.

The ESO Data Reduction Pipeline Version\,1.3 \cite{Jung_VLT} and the
QFitsView\footnote{Written by Thomas Ott (MPE, Garching);
  http://www.mpe.mpg.de/$\sim$ott/QFitsView} software are used for the
data analysis.  In the first steps, the images are cleaned from bad
pixels, flat-fielded and wavelength calibrated.  A distortion
correction is then applied and the FoV of SINFONI is reconstructed.
The output of this procedure is a three-dimensional (3D) cube
containing about 2000 images of the source in 0.5\,nm wavelength
steps.  

Averaging the cube along the wavelength axis enables the effective
construction of a broad-band image. H-band and K-band images were thus
extracted from PG\,1553+113 data. These images are spatially
unresolved and no underlying host galaxy is detected.
 
A spectrum is extracted from the final 3D cube by summing the individual
spectra within an aperture of 5-pixel radius centered on the brightest
peak in the image. Observations 
of two G2V standard stars (HIP071527, HIP088235),
near both in time and air-mass to the target exposures, are
used to correct for strong atmospheric (telluric) absorption.  By
dividing the science target spectrum by the standard star spectrum,
the atmospheric absorption effects are minimized (see, e.g.,
\cite{Maiolino}, \cite{Vacca}). However, this step introduces features
into the resulting spectrum that are intrinsic to the standard star.
Here, these features are accounted for by multiplication of the
normalized high-spectral-resolution atmospheric-transmission-corrected
solar spectrum\footnote{NSO/Kitt Peak FTS data of the sun used here
were produced by NSF/NOAO.}  that has been degraded to the SINFONI
spectral resolution.

Flux calibration at infrared wavelengths is somewhat more problematic
than at optical wavelengths since no standard spectral catalog exists.
A theoretical continuum spectrum of a G2V star could in principle be
used to correct for the throughput of the spectrograph.  However, the
flux distribution of such stars, as well as later-type ones, in the
infrared is not well-reproduced by theoretical modeling, and a small
error in the stellar parameters (e.g., the effective temperature or
the metallicity) can yield a large error in the continuum.  Instead
the wavelength-dependent flux calibration is determined from
observations of two intrinsically featureless (apart from strong
hydrogen lines) B stars (HIP088201, HIP085812).  Here the hydrogen
lines are manually fit and removed from the B-star spectrum, and the
result is then divided by a blackbody spectrum at the effective
temperature of the stars (18500 K and 11900 K, respectively) 
to correct for the continuum slope of each
standard star. As the H and K bands are in the Rayleigh-Jeans regime
of the blackbody spectrum, the steepness of the continuum slope for
these B stars is not particularly sensitive to small errors in the
effective temperature. The final step of the flux calibration is
setting the absolute flux scale. IR aperture photometry of
PG\,1553+113 was already performed but the source is known to be
variable. Therefore 2MASS H magnitudes of the standard stars,
converted to flux density, are used instead to scale the total flux
observed here. This value, 12.7$\pm$1.3 mJy at 1.65$\mu$m, is
approximately half of that observed by \cite{Bersanelli} and is
comparable with R-band fluxes observed (\cite{MAGIC_1553}) in March
2006. It should be noted that the latter (March 15, 2006) of the
near-IR observations reported here were performed two days before an
outburst in the R-band reported by \cite{MAGIC_1553}.

   \begin{figure*}
   \centering
      \includegraphics[width=0.95\textwidth]{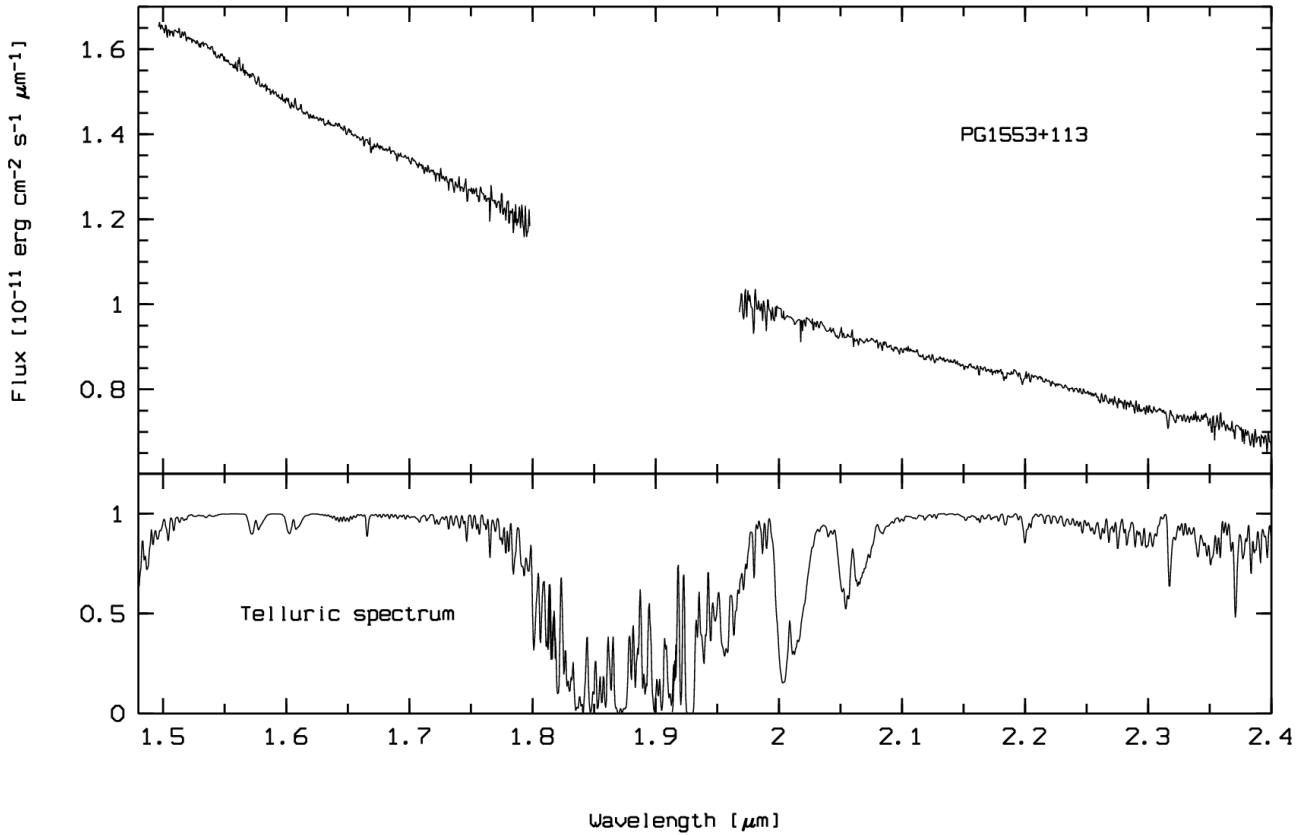}
      \caption{Top: H+K-band spectrum of PG\,1553+113 extracted 
from a 5-pixel radius aperture. The gap is due to the highly 
reduced atmospheric transmission between H and K bands. 
Bottom: Telluric spectrum extracted from the standard stars.}
         \label{IR_spectrum}
   \end{figure*}

   The final H+K-band spectrum of PG\,1553+113 is shown in
   Figure~\ref{IR_spectrum}. The signal-to-noise ratio is $\sim$250 in
   the H-band and $\sim$70 in the K-band.  These measurements 
   are the most sensitive ever
   recorded in the H- and K-bands, and are comparable
   to the most sensitive spectroscopy of PG\,1553+113
   at other wavelengths (see, e.g,
   \cite{Carangelo_03,no_lines}).  The lack of data between
   $\sim$1.80$-$1.95 $\mu$m is a result of highly-reduced atmospheric
   transmission between the H and K bands. The observed near-IR
   spectrum is featureless apart from some residuals from the
   atmospheric correction. Thus, in neither the broad-band images nor
   in the spectrum are the influences of the gas of a host galaxy or
   the AGN detected, even though PG\,1553+113 is bright in the IR.  As
   a result, a redshift determination from these observations is not
   possible.

\section{Discussion}

The presence of Suzaku X-ray observations, 
simultaneous with the HESS measurements, greatly improves
the possibilities for accurately modeling the underlying
physics in PG\,1553+113.   However, the observed VHE spectrum is
very soft ($\Gamma=4.46\pm0.34$), which could be in 
large part due to the redshift-dependent absorption of 
VHE photons on the EBL.  Modeling the
VHE portion of the SED of PG\,1553+113 requires the effects of this
absorption to be removed.  Unfortunately the redshift of the object is
still unknown, despite the very sensitive new measurements presented
in this article.  Therefore, modeling of the VHE emission from
PG\,1553+113 is extremely difficult given the large range of possible
intrinsic spectra.

PG\,1553+113 is not the only bright
BL\,Lac for which the underlying host galaxy is not detected, even though
deep, high signal-to-noise ratio observations were conducted (e.g., O'Dowd
\& Urry 2005, \cite{no_lines}) for many. The properties of their 
host galaxy or their AGN may be such that it is 
presently not possible to derive
their intrinsic nature and hence redshift. In these cases, it is
possible that the properties of the host are not those of giant
elliptical galaxies, but correspond to the faint end of the elliptical
luminosity distribution (M$_R$ $\sim$ $-$20) or that the AGN may even be
hosted by a dwarf galaxy, making these BL\,Lac objects the radio-loud
equivalent to the low-mass black-hole Seyfert galaxies \cite{dwarf}. 
If this is the case, it will be very difficult to derive a firm 
redshift for PG\,1553+113.

As mentioned in the introduction, it is
possible to derive upper limits on the object's redshift using VHE
measurements, some basic assumptions about the source's intrinsic
spectrum, and an EBL density model.  As yet, none of the limits derived
in this fashion are particularly constraining.  The strongest, and
also based on the most assumptions, is $z<0.42$ \cite{Mazin_limit}.
Following exactly the same methodology as in Aharonian et al. (2006a),
the VHE spectrum measured here from PG\,1553+113 limits the redshift to
$z<0.69$.

\section{Conclusion}

With a data set that is $\sim$3 times larger than previously
published (\cite{HESS_discovery}), the HESS signal from PG\,1553+113 is 
now highly significant ($\sim$10$\sigma$).  
Thus, the evidence for VHE emission previously reported
is clearly verified.  However, the flux measured in 2005
is now $\sim$3 times higher than first reported
due to an improved calibration of the absolute energy scale
of HESS. The statistical error on the VHE photon index is now
reduced from $\sim$0.6 to $\sim$0.3.  Nevertheless,
the error of 0.34 is still rather large, primarily due
to the extreme softness of the observed spectrum ($\Gamma=4.46$).  
The total HESS exposure on PG\,1553+113 is $\sim$25 hours.  Barring
a flaring episode, not yet seen in two years of observations,
a considerably larger total exposure ($\sim$100 hours) would be
required to significantly improve the spectral measurement.  This large
exposure is unlikely to be quickly achieved.  However,
the VHE flux from other AGN is known to vary dramatically
and even a factor of a few would reduce the observation
requirement considerably. Should such a VHE flare occur, 
not only will the error on the measured VHE spectrum be 
smaller, but the measured photon index may 
also be harder (see, e.g., \cite{VHE_hardening}).  Both effects would 
dramatically improve the redshift constraints and correspondingly
the accuracy of the source modeling. Therefore, the VHE flux
from PG\,1553$+$113 will continue to be monitored by HESS.
In addition, the soft VHE spectrum makes it an ideal target for
the lower-threshold HESS Phase-II \cite{HESSII} which should
make its first observations in 2009.

\begin{acknowledgements}

The support of the Namibian authorities and of the University of Namibia
in facilitating the construction and operation of H.E.S.S. is gratefully
acknowledged, as is the support by the German Ministry for Education and
Research (BMBF), the Max Planck Society, the French Ministry for Research,
the CNRS-IN2P3 and the Astroparticle Interdisciplinary Programme of the
CNRS, the U.K. Science and Technology Facilities Council (STFC),
the IPNP of the Charles University, the Polish Ministry of Science and 
Higher Education, the South African Department of
Science and Technology and National Research Foundation, and by the
University of Namibia. We appreciate the excellent work of the technical
support staff in Berlin, Durham, Hamburg, Heidelberg, Palaiseau, Paris,
Saclay, and in Namibia in the construction and operation of the
equipment. Based on ESO-VLT SINFONI program 276.B-5036 observations.

\end{acknowledgements}

\end{document}